# Stability Analysis of Superconductivity in *P*6/*mmm*-$LaSc_2H_{24}$ and its Experimental Reproducibility from La-Sc Alloys

Dmitrii V. Semenok [1,†,*], Ivan A. Troyan [2,†,*], Zhongyan Wu [1,†], Di Zhou [1], Emil A. Yuzbashyan [3], Boris L. Altshuler [4,*], and Viktor V. Struzhkin[2,5]

[1] Center for High Pressure Science & Technology Advanced Research, Bldg. #8E, ZPark, 10 Xibeiwang East Rd, Haidian District, Beijing, 100193, China

[2] Shanghai Key Laboratory of Material Frontiers Research in Extreme Environments (MFree), Shanghai Advanced Research in Physical Sciences (SHARPS), 68 Huatuo Rd, Bldg #3 Pudong, Shanghai 201203, China

[3] Department of Physics and Astronomy, Center for Materials Theory, Rutgers University, Piscataway, 08854, NJ, USA

[4] Physics Department, Columbia University, 538 West 120th Street, New York, 10027, NY, USA

[5] Center for High Pressure Science & Technology Advanced Research, 1690 Cailun Rd, Bldg 6, Pudong, Shanghai 201203, China

[†]These authors contributed equally to this work

Corresponding authors: Dmitrii V. Semenok, Ivan A. Troyan, and Boris L. Altshuler

**Email:** dmitrii.semenok@hpstarac.cn , itrojan@sharps.ac.cn , altshulerboris1@gmail.com

## Abstract

In this work, we analyze the feasibility of room-temperature superconductivity in the lanthanum-scandium hydride *P*6/*mmm*-$LaSc_2H_{24}$. We demonstrate that the electron-phonon coupling calculations performed using the σ-broadening of the double δ-function at the Fermi surface lead to a very strong dependence of $T_c(\sigma)$ on the arbitrary σ, whereas the tetrahedral method for the electron-phonon interaction is free from this drawback and gives $T_c$ > 300 K at 300 GPa in agreement with previous predictions. By analyzing the stability of metallic state of $LaSc_2H_{24}$ at 250-300 GPa, we show that this compound is at the edge of the stability region ($\xi$ = 0.54), similar to *fcc* $LaH_{10}$ at 140-150 GPa. Seven experimental attempts to synthesize $LaSc_2H_{24}$ at 250-280 GPa starting from the (La, $Sc_2$) alloy are unsuccessful and indicate the absence of even traces of superconductivity at 245-300 K in all the resulted La-Sc-H hydrides. The method for preparing the precursor by simultaneous deposition of La and Sc metals may be a key factor for the successful synthesis of $LaSc_2H_{24}$.

## Introduction

The superconducting critical temperature of metal hydrides has increased substantially over the past five years, reaching values up to 280 K (+7 °C) at pressures of about 150–170 GPa in several experiments on La-based polyhydrides (1-6). Partial resistive transitions at 260–280 K were already observed in the first study of superconductivity in $LaH_{10}$,(1) and were later confirmed by AC susceptibility measurements (3). In 2024, optical pump-probe ultrafast spectroscopy and resistive measurements confirmed the formation of unknown lanthanum hydrides with a higher $T_c \approx 265$ K (4) compared to thermodynamically stable $LaH_{10}$ ($T_c$ = 250 K) (1, 2). Finally, in 2025–2026, experiments with the La-Sc-H system revealed that lanthanum-scandium hydrides contain a phase that exhibits a resistivity drop already at 274 K. The onset of the possible superconducting transition occurs already at ≈280 K (6) in the radio-frequency contactless measurements (7).

At the same time, room-temperature superconductivity ($T_c$ > 300 K at 250 GPa) was theoretically predicted in thermodynamically stable $LaSc_2H_{24}$ (8), motivating us to explore the pressure range above 2 Mbar. Recent $^1$H NMR and the radio-frequency (RF) study of two-phase mixture of $LaH_{12}$ revealed a pronounced RF transmission anomaly and a sharp increase in the NMR relaxation time in the sample around 270–280 K at a pressure of 165 GPa (9). This anomaly may have a superconducting origin and may be responsible for the sharp drops in electrical resistance previously observed in La-based polyhydrides.

The results of the experimental synthesis and measurements of the superconducting properties of *P6/mmm*-$LaSc_2H_{24}$ were presented for the first time in 2025 (10). The authors of Refs. (10, 11) indicated that this compound can be obtained via laser heating of a La/Sc mixture or La-Sc alloys at about 260 GPa and that it exhibits room-temperature superconductivity with an onset $T_c$ of up to 298 K (11). According to the authors of Ref. (11), the theoretical basis for this experimental result is the *P*6/*mmm*-$LaSc_2H_{24}$ model proposed in Ref. (8).

In this work, we investigate how different σ-broadening values affect theoretical $T_c(LaSc_2H_{24})$ at different pressures, study the stability of the metallic state of $LaSc_2H_{24}$ using the recently developed electronic heat capacity criterion ($C_{el} \geq 0$), calculate Eliashberg functions using the tetrahedral method for a series of isostructural hexagonal hydrides: *P6/mmm*-$La_3H_{24}$, $Sc_3H_{24}$, and $LaSc_2H_{24}$ at 300 GPa, and describe seven unsuccessful attempts to reproduce room-temperature superconductivity in the La-Sc-H system using ($La,Sc_2$) alloys of the La:Sc 1:2 composition as starting precursors for the synthesis at 254-280 GPa.

## Results

### I. Theoretical analysis

We previously developed a framework for describing the instability of electron-phonon superconductors, particularly hydrides, at strong electron-phonon coupling. In this framework, the metallic state becomes unstable toward a reconstruction of the hydrogen sublattice, which may lower the crystal symmetry, lead to partial hydrogen loss, and open a gap or a pronounced depression in the electronic density of states near the Fermi level, replacing the superconducting gap (12-14). The key quantity is the stability parameter ξ,

$$\xi \equiv \max_{T}\left\{\int_0^\infty g\left(\frac{\omega}{2\pi T}\right)\frac{2\alpha^2F(\omega)}{\omega}d\omega\right\} < 1, \qquad (1)$$

where $g(x) = 6x^2 + 12x^3 Im\,\psi'(ix) + 6x^4 Re\,\psi''(ix)$, and $\psi(x)$ is the digamma function (12). The condition $\xi < 1$ is the absolute stability condition: at $\xi = 1$ the electronic specific heat vanishes, and for $\xi > 1$ the metallic state is unstable even with respect to infinitesimal deviations from thermal equilibrium. The actual phase transition, however, is expected to occur earlier, at a critical value $\xi_c < 1$, because it is a first-order transition. The value of $\xi_c$ is not fixed by the general theory and may depend on the crystal structure, electronic band structure, carrier density, and other material-specific details.

A comparison with available data for hydrides suggests that this instability occurs at $\xi_c \approx 0.5$–$0.55$. For example, among known superhydrides, *fcc*-$LaH_{10}$ below 175 GPa is the only case for which ξ, calculated directly from anharmonic Eliashberg functions (15), exceeds 0.5 and likely approaches 0.55 near the verge of structural stability around 140 GPa (12, 16). Below this pressure, the cubic structure of $LaH_{10}$ is already strongly distorted, for example through an $Fm\overline{3}m \rightarrow C2/m$ transformation. At the same time, $T_c$ in $LaH_{10}$ remains nearly constant, $T_c$ = 240–250 K, over the pressure range 140–180 GPa (16).

Analysis of the Eliashberg functions of several hydrides near their stability limit (12) suggests a possible microscopic route by which strong electron–phonon coupling destabilizes the metallic hydrogen sublattice. Upon decompression, phonon softening and anharmonic effects enhance the contribution of low-frequency hydrogen modes to the stability parameter ξ until it reaches the critical value $\xi_c$. At the microscopic level, this process may involve an unstable intermediate state, resembling a hydrogen quasi-molecule near the limit of its stability, which is disrupted by the field of the heavy-atom sublattice *M* (Fig. 1C-D). Conversely, a depletion of electron density in the antibonding orbitals of $H_2$, for example after hole capture $h^+$, favors the shortening of the H–H distance and the formation of a stable $H_2$ molecule. Such quasi-molecular hydrogen configurations have recently been studied theoretically in $ScH_{12}$ and $MgH_{12}$ (17). The lifetime of this intermediate state should be of order $\tau \sim 1/\omega_{min}$, where $\omega_{min}$ is the minimum frequency of the relevant hydrogen-sublattice mode. This provides a possible route for replacing atomic hydrogen by molecular hydrogen and for the accumulation of molecular defects (Fig. 1D) in the metallic hydrogen sublattice.

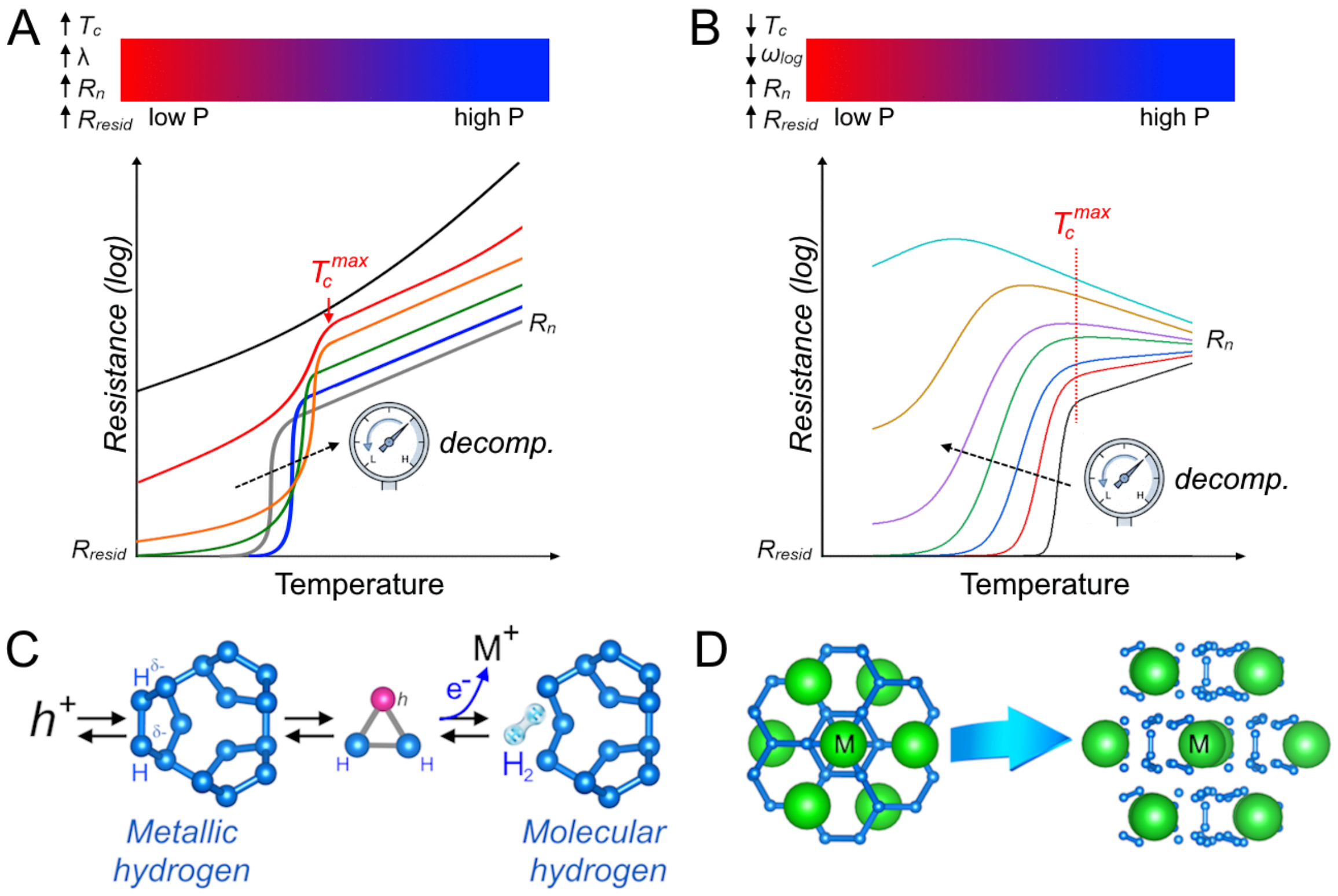

**Fig. 1.** Qualitative mechanism of possible instability of the metallic hydrogen sublattice in superhydrides during decompression. (A) Experimental picture of hydride decompression with an increase in the critical temperature, sample resistance, electron-phonon interaction parameter, and, eventually, disappearance of superconductivity. The color scale corresponds to the transition from low (blue) to high (red) values of normal state and residual resistance, $T_c$, λ upon going from high pressures (high P) to low pressures (low P). The dominant contribution to $T_c$ is made by the increase in λ during decompression. (B) Another possible scenario for superhydride decomposition during decompression, characterized by the dominant contribution to $T_c$ coming from the decrease in $\omega_{log}$. This pathway is realized, for example, for sulfur hydrides (18). (C) Potentially reversible mechanism of formation of a hydrogen quasi molecule inside of the metallic hydrogen sublattice due to fluctuations of the local electron density. "$h^+$" denotes a hole. (D) Further decay of the metallic hydrogen sublattice during decompression with the formation of numerous $H_2$ molecules and a decrease in the symmetry of the heavy atom sublattice (M).

Experimentally, these scenarios can be followed during decompression of superhydrides. At first, the superconducting critical temperature $T_c$ increases as pressure is lowered, while the normal-state resistance $R_n$ also increases because of stronger electron scattering by phonons (Fig. 1A). Alternatively, in the case of a dominant effect of the decrease in $\omega_{log}$ on $T_c$, the latter decreases during decompression (Fig. 1B). In any case, as decomposition progresses, a nonzero residual resistance $R_{resid}$ appears, likely due to the formation of nonsuperconducting regions at grain boundaries. Upon further decompression, $R_{resid}$ grows and eventually masks the superconducting transition of what has become "island" superconductivity. In this way, a hydride with an atomic hydrogen sublattice evolves into a nonsuperconducting polyhydride with a molecular hydrogen sublattice. The symmetry of the metal sublattice does not necessarily change strongly, even at much lower pressures of order 10–20 GPa, as recently demonstrated for $LaH_{12}$ (9) and $La_4H_{23}$ (19-21) In such cases, one may observe mainly a broadening of the diffraction peaks associated with the metal sublattice. Other hydride phases with lower symmetry, lower $T_c$, reduced density of states, and lower hydrogen content can also form simultaneously along this metal-to-metal transition pathway.

Turning to $P6/mmm$-$LaSc_2H_{24}$, we find that even within the harmonic approximation at 290–300 GPa the stability parameter reaches $\xi_{max} \approx 0.55$ (Fig. 2A). This value is at the upper end, or slightly above, the empirical range $\xi_c \approx 0.5$–0.55 inferred from known hydrides near their stability limit. Since $\xi_c$ is not a universal theoretical constant, this observation does not by itself prove that the metallic state of $LaSc_2H_{24}$ is unstable. It does, however, suggest that this phase may lie close to the stability boundary, and that the stability of its metallic state requires a more detailed analysis, including anharmonic effects.

What is the nature of room-temperature superconductivity in $P6/mmm$-$LaSc_2H_{24}$, taking into account the anharmonicity of hydrogen sublattice vibrations? As an analysis of the Eliashberg functions provided by the authors of Ref. (8) showed, the $T_c$ of this La-Sc hydride is very dependent on the broadening parameter (σ), which is used to sum the double delta function in formula (2) within the Quantum Espresso code (Fig. 2b) (22, 23)

$$\alpha^2 F(\omega) = \frac{1}{N(\epsilon_F)} \sum_{\mathrm{mnq\nu}} \delta(\omega - \omega_{\mathrm{q\nu}}) \sum_{\mathrm{k}} |g_{\mathrm{mn\nu}}(\mathrm{k}, \mathrm{q})|^2 \times \delta(\epsilon_{\mathrm{m,k+q}} - \epsilon_{\mathrm{F}}) \delta(\epsilon_{\mathrm{n,k}} - \epsilon_{\mathrm{F}}), \quad (2)$$

where delta functions are approximated as

$$\delta(\epsilon_{n,k} - \varepsilon_{\mathrm{F}}) \approx \frac{1}{\sigma\sqrt{2\pi}} \exp\left[-\frac{(\epsilon_{n,k} - \varepsilon_{\mathrm{F}})^2}{2\sigma^2}\right]. \quad (3)$$

Indeed, due to the peculiarities of the electronic structure and the presence of peaks in the density of electronic states $N(\varepsilon)$ near the Fermi energy $\varepsilon_F$ in $LaSc_2H_{24}$ (8), increasing σ from 0.005 Ry to 0.05 Ry reduces $T_c$ by 50-60 K, moving this compound from the class of room-temperature superconductors (Fig. 2B). The choice of the "correct" σ can be made based on the experiment. It is important to emphasize once again that $LaSc_2H_{24}$ is a room temperature superconductor only at $\sigma \leq 0.015$ Ry.

We can approach the problem in a different way. Without addressing the question of which σ is the most correct, we can analyze the stability function $\xi(\sigma, P)$ for $LaSc_2H_{24}$ (Fig. 2C). Using the anharmonic Eliashberg functions provided by authors of Ref. (8), we found that at 150-167 GPa the metallic state in this compound is in the "red" zone ($\xi > 0.55$, Fig. 2C) and is likely unstable at all studied σ values, while at 200 GPa it is stable only at large broadening $\sigma > 0.035$ Ry, where $T_c(LaSc_2H_{24}) < 250$ K. Thus, the room-temperature superconductivity in $LaSc_2H_{24}$, for example, at 167 GPa (8), cannot be realized due to the instability of the metallic hydrogen sublattice, and the need to use of the highest σ values for its formal stabilization. Indeed, the first stage of superhydride decomposition is structural disruption, the appearance of defects in the H-sublattice, and the symmetry breaking of the heavy-atom sublattice, with a corresponding broadening of the electron band structure and a decrease in $T_c$ (12).

Eventually, at higher pressures of 250-300 GPa, the metallic state of $LaSc_2H_{24}$ is stable at all σ, but this comes at the cost of a significant decrease in $T_c$ below 280 K (Fig. 2D). Thus, we conclude that, regardless of the σ selection criterion, $LaSc_2H_{24}$ can be a near-room-temperature superconductor only at pressure above 2 Mbar. The maximum $T_c$ can reach 270-280 K (Fig. 2D) given the fact that the solutions of the Eliashberg equations for $T_c$ are somewhat higher than what is predicted by the rather conservative Allen-Dynes formula (24, 25).

Another approach to calculating $T_c$ without using arbitrary σ-broadening is the tetrahedral method (26), in which we represent $\epsilon_{\mathrm{n,k}} = \sum_{j=1}^{4} x_{j,k} \epsilon_{\mathrm{n}}$ as a linear function of specially introduced coordinates ($x_{j,k}$), measured from the vertices of tetrahedra constructed on a grid shifted relative to the Γ-point of BZ, so that none of the tetrahedra's vertices fall within the Γ-point. Then, in formula (2), the sum over $\bar{k}$ can be replaced by the integral over the BZ, and, further, by the sum of volumes of tetrahedra's parts. The tetrahedral method has been used to study many conventional superconductors, such as sulfur hydrides (27), silicon (28), $YNi_2B_2C$ (29), $Li_{2x}BC_3$ (30), and many others.

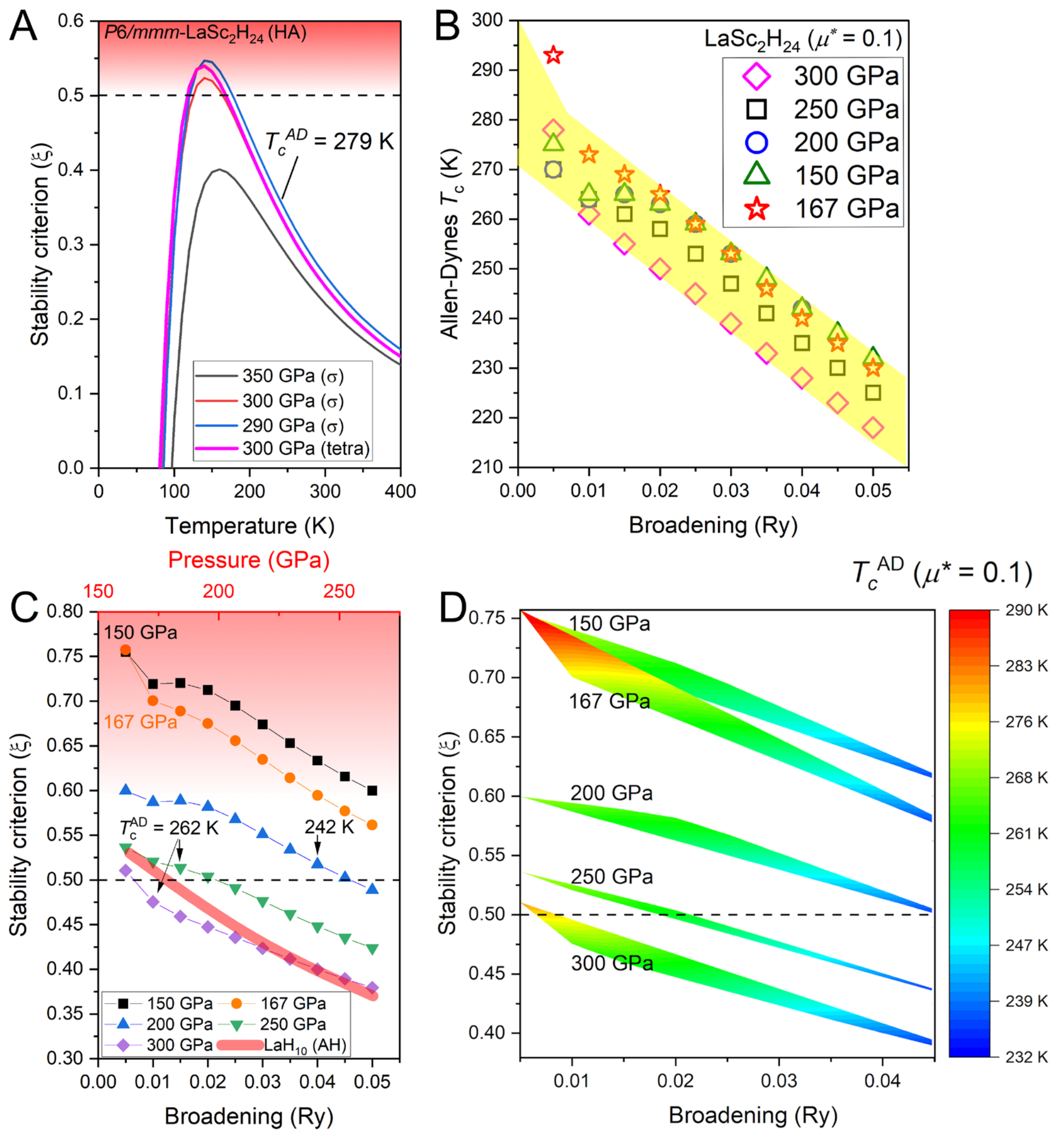

**Fig. 2.** Metallic state stability and superconducting properties of *P6/mmm*-$LaSc_2H_{24}$ under pressure. (A) Temperature-dependent stability criterion ξ(T) for $LaSc_2H_{24}$ in harmonic approximation (HA) at 290-350 GPa, showing possible violation of the stability criterion (ξ > 0.5) with the maximum $\xi_{max}$ = 0.55 at 290 GPa. "σ" corresponds to calculations using formula (3), and "tetra" corresponds to the tetrahedral method of calculating the electron-phonon interaction. Eliashberg functions for the calculations were taken from Ref. (8) as is. (B) Critical temperature $T_c$ as a function of broadening parameter σ at different pressures (150-300 GPa), demonstrating strong sensitivity to σ with $T_c$ decreasing by 50-60 K as σ increases from 0.005 to 0.05 Ry, potentially eliminating the room-temperature superconductivity. Data is taken from Ref. (8) (c) Stability criterion ξ versus broadening σ at various pressures, revealing three stability regions: red zone (150-167 GPa) where metallic state is unstable at all σ values; intermediate zone (200 GPa) where stability requires σ > 0.035 Ry but $T_c$(AD) < 250 K; and green zone (250-300 GPa) with a stable metallic state at all σ but reduced $T_c$. The upper red pressure scale corresponds to the data for $LaH_{10}$. (D) Phase diagram mapping stability criterion ξ versus broadening σ at different pressures (150-300 GPa) with color-coded $T_c$ values, demonstrating that a stable metallic state with maximum $T_c$ (AD) ≈ 270 K ("green" region of the color bar) is only achievable at high pressures (250-300 GPa), which makes the realization of the room-temperature superconductivity in $LaSc_2H_{24}$ unlikely.

Calculations employing the tetrahedral method for electron-phonon interaction in the *P6/mmm*-$La_3H_{24}$, isostructural to $LaSc_2H_{24}$, result in a harmonic $T_c$ of less than 200 K at 300 GPa (Fig. 3A). This outcome is consistent with the observed trend in which $T_c(XH_8)$ is lower than $T_c(XH_{9\text{-}10})$ for binary superhydrides. For the hexagonal $LaH_{9\text{-}10}$, the $T_c$ ranges from 160 to 215-220 K (5, 31-33). Similarly, the calculated value of $T_c(Sc_3H_{24})$ = 223 K at 300 GPa (Fig. 3B, via the Allen-Dynes formula (AD) (25)) is significantly higher than previously obtained experimental $T_c$'s in scandium hydrides, which likely have the formula $ScH_{3\text{-}4}$ (34). It would seem that the combination of two poor superconductors, $LaH_8$ and $ScH_8$, cannot lead to a significant increase in superconducting properties in La-Sc-H ternary hydrides, but this turns out to be incorrect: direct calculations at 300 GPa indicate a sharp increase in the electron–phonon interaction amplitude, and give $T_c$(AD) up to 307 K (Fig. 3E, $\mu^* = 0.1$). The electron–phonon coupling parameter exceeds the maximum value for the Einstein phonon spectrum $\lambda = 4.77 > \lambda_{max} = 3.69$ (12). This indicates an exceptionally strong electron–phonon interaction in $LaSc_2H_{24}$, which may go beyond the scope of the Migdal–Eliashberg theory and requires, at a minimum, taking into account anharmonic effects. Despite the unrealistically large value of λ, the stability parameter is within the stability range and only slightly exceeds ½: ξ = 0.54.

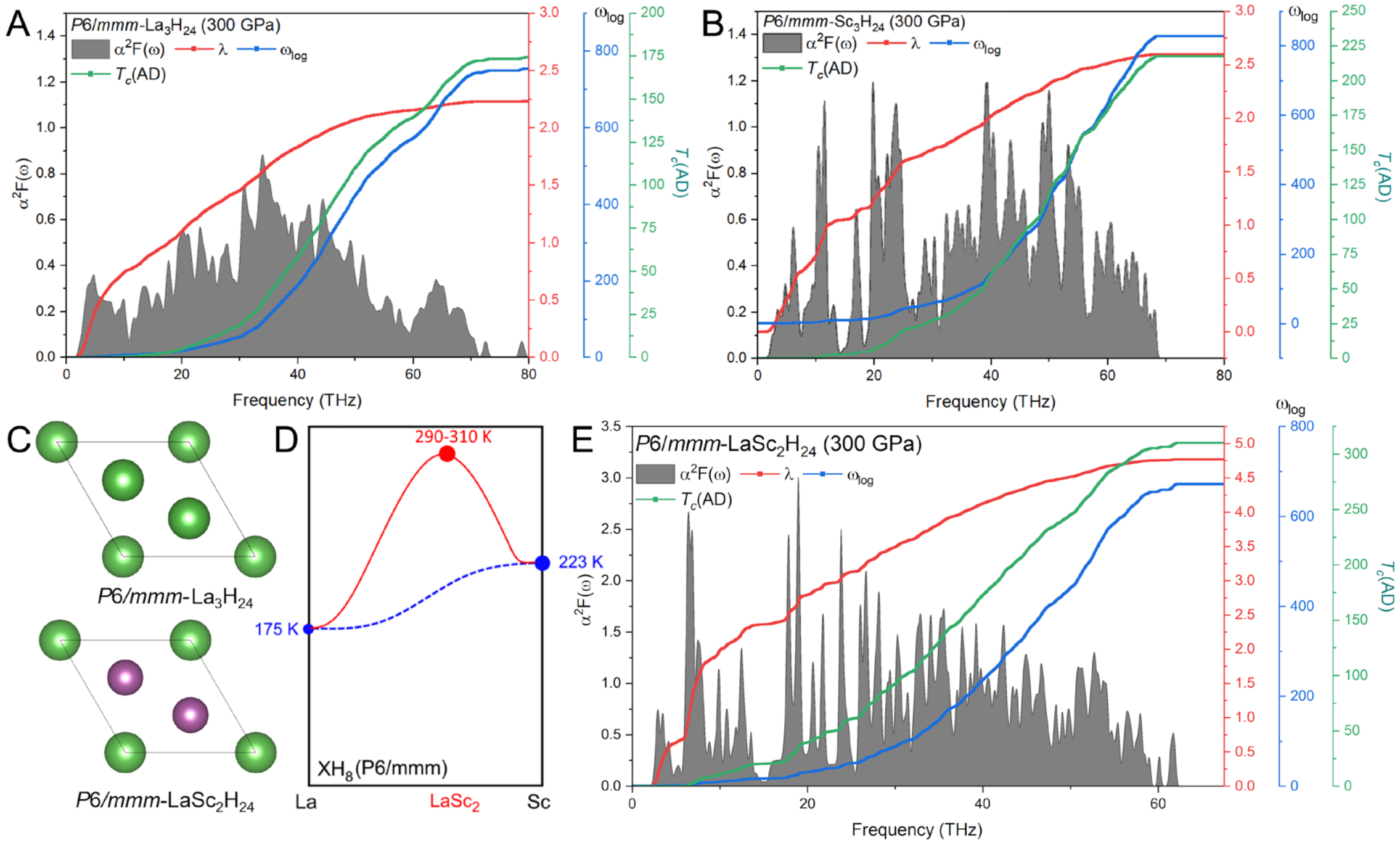

**Fig. 3.** Harmonic Eliashberg functions and superconducting state parameters for isostructural (A) *P*6/*mmm*-$La_3H_{24}$, (B) $Sc_3H_{24}$, and (E) $LaSc_2H_{24}$ at 300 GPa in harmonic approximation. The tetrahedral method was used for Quantum Espresso calculations with a k-mesh of 12×12×12, a q-mesh of 4×4×4, and for the electron band structure calculations, we applied a denser k-mesh of 24×24×24. $T_c$ is estimated by the Allen-Dynes formula (25) with $\mu^* = 0.1$, other parameters (λ, $\omega_{log}$) obtained using a previously developed postprocessing scripts (35). (C) Structures of $X_3H_{24}$, where X = La (green), Sc (violet), and $LaSc_2H_{24}$. (D) The expected behavior of the $T_c(x)$ function for the La-Sc-H system (blue dashed line, where $x$ is the scandium concentration), as observed in disordered alloys. The red line qualitatively corresponds to the expected behavior of true ternary hydrides, including $LaSc_2H_{24}$.

The appearance of a sharp maximum in the $T_c(x)$ vs Sc concentration curve at $x = 0.66$ in the La-Sc-H system (Fig. 3D) is unexpected for at least two reasons:

- It has been established that cubic hydrides generally exhibit higher $T_c$ values in comparison to hexagonal ones. Examples include *hcp*-$CeH_9$ / *fcc*-$CeH_{10}$ (36), *hcp*-$ThH_9$ / *fcc*-$ThH_{10}$ (37), and *hcp*-$LaH_{9-10}$ / *fcc*-$LaH_{10}$ (2, 31). On this basis, we can conclude that if cubic ternary hydrides exist in the La-Sc-H system, then their $T_c$ can be even higher than that of $LaSc_2H_{24}$.

- A specific proportionality exists between the $T_c$ and hydrogen content, and for hydrides of formula $XH_n$ for $n$ = 3…10, this relationship is monotonic. For instance, the following sequence has been established: $T_c(LaH_3) < T_c(LaH_4) < T_c(La_4H_{23}) < T_c(LaH_6) < T_c(LaH_{9-10}) < T_c(LaH_{10})$ (38). A similar observation has been made in the case of yttrium (39-41), cerium (36), thorium (37), and calcium (42, 43). However, the synthesis of scandium hydrides, which exhibit the maximum transition temperature ($T_c$) only about 70 K under pressure of 180 GPa (44), poses significant challenges. It follows that increasing the hydrogen content to $LaSc_2H_{27}$ or $LaSc_2H_{30}$ should lead to an increase in the critical temperature of superconductivity.

Finally, it is worth addressing the applicability of Anderson's theorem to the *P*6/*mmm*-$LaSc_2H_{24}$. Consider a scenario in which the La–Sc precursor alloy exhibits a compositional deviation of about 10% from stoichiometry, corresponding to a range spanning from $La_{0.4}Sc_{0.6}$ to $La_{0.25}Sc_{0.75}$. In such a case, excess Sc atoms would partially occupy La Wyckoff positions in a disordered fashion, while a La excess would result in a random substitution at Sc crystallographic sites within the *P*6/*mmm* structure. Nevertheless, in accordance with Anderson's theorem (6, 45), such non-magnetic isoelectronic disorder introduced by random site substitution is not expected to significantly change $T_c$. Consequently, precise control over the La:Sc stoichiometric ratio of 1:2 in the precursor alloy should not constitute a critical synthesis parameter for achieving superconductivity in this compound.

## II. Experimental attempts to synthesize $LaSc_2H_{24}$ from La-Sc alloys

The first experiment was performed using an electric DAC SL-1 (Fig. 4). The La-Sc alloy was prepared by arc melting a sample of La and Sc weighing approximately 1–2 g for 1–3 minutes at a temperature above 2000 °C in an argon atmosphere. Additional mixing of the elements was observed during laser heating of samples already inside the high-pressure DAC. The diamond anvils used in the experiment had a culet diameter of 30 μm and were protected by a sputtered submicron aluminum oxide layer. A composite gasket made of W/$CaF_2$/epoxy was prepared for this experiment.

A pre-compressed La:Sc alloy particle (1:2 ratio) was studied using an electron microscope and the Energy Dispersive X-ray (EDX) analysis (SI Appendix, Fig. S1). A homogeneous region of approximately 10 × 10 μm² with an elemental ratio of 1:2 (SI Appendix, Fig. S1B) was selected for the loading with ammonia borane ($NH_3BH_3$). A series of independent experiments with DACs Z1-Z4 was performed using homogeneous $LaSc_2$ alloy prepared by vacuum melt spinning from an arc-melted precursor. Centimeter-sized fragments with the desired La:Sc ratio of 1:2 (SI Appendix, Fig. S2-S3) were selected from the resulting ribbons. X-ray diffraction of the starting material shows the presence of a mixture of *dhcp*-La, *fcc*-La and *hcp*-Sc phases, typical for these metals at ambient pressure (SI Appendix, Fig. S4).

The initial pressure before laser heating was 280 GPa (Akahama scale (46)). After multiple pulsed IR laser heating (1.06 μm laser, 0.05 s pulse length), the pressure in DAC SL-1 dropped to 264 GPa (Fig. 4D), and pronounced changes in the morphology and geometry of the sample were observed (Fig. 4A–C). Part of the sample melted and was pushed outside the culet during the laser heating process.

Subsequent electrical resistance measurements were performed over the temperature range of 250–300 K and demonstrated normal-state metallic behavior (Fig. 4E, F). The resistance curves from different contact channels showed generally consistent behavior, confirming sample continuity and homogeneity across the probed area. No signs of a possible high-temperature superconducting transition between 242 K and 295 K were observed. This DAC failed without any external influence within a few days after the resistance measurements were completed.

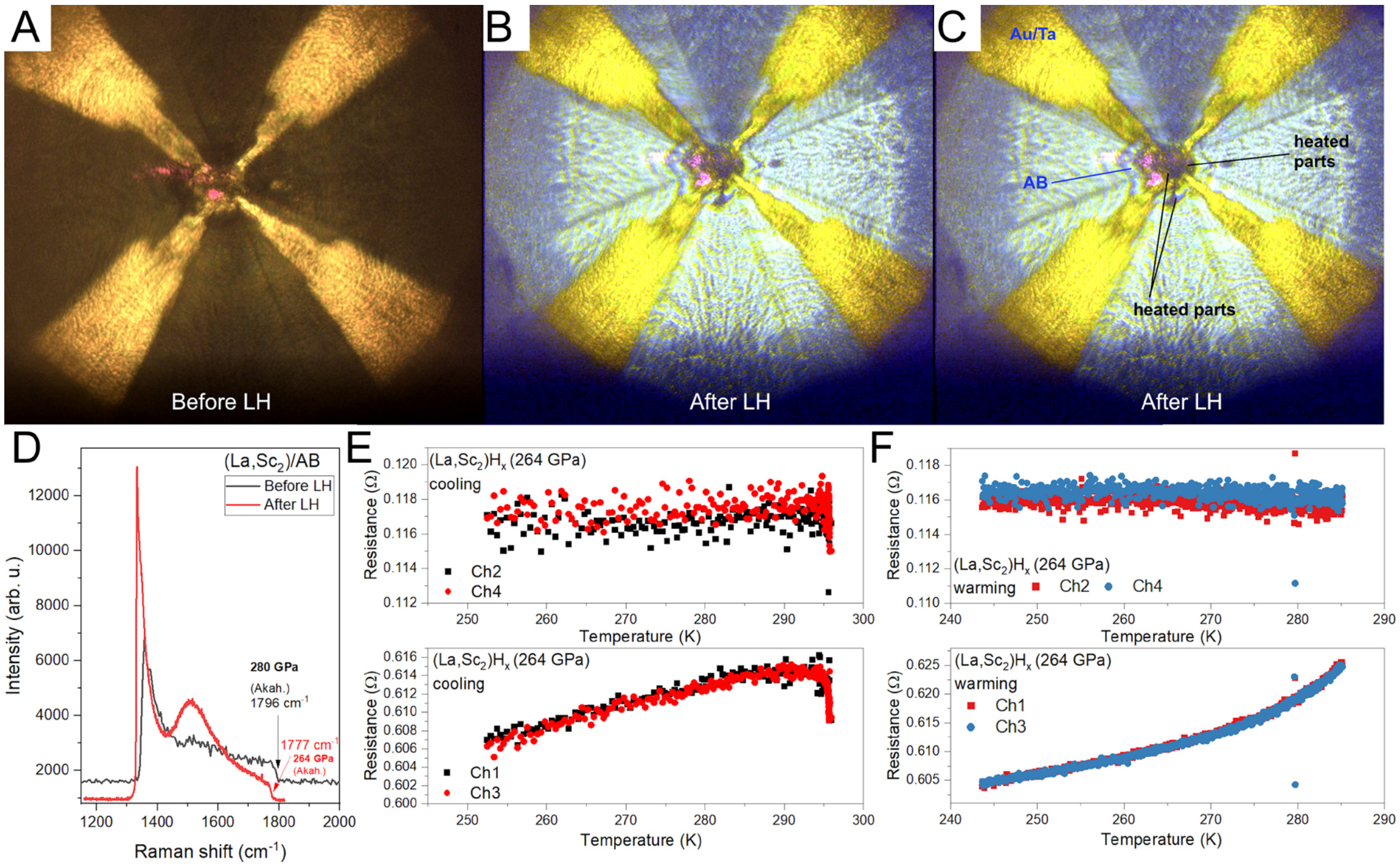

**Fig. 4.** Synthesis of (La,Sc)$H_x$ using infrared pulsed laser heating (LH) under high pressure in DAC SL-1. (A-C) Optical micrographs showing the sample chamber and electrical contacts before and after laser heating at high pressure. (A) Initial picture of the (La,Sc)/AB (ammonia borane) sample before LH. (B) State after LH showing a blackening of the sample's surface. (C) View of sample after LH, highlighting the regions that experienced high temperatures (labeled "heated parts"), the Au/Ta electrodes, and AB/$CaF_2$ insulating gasket. (D) Raman spectra collected before and after laser synthesis. During laser heating, pressure in the DAC decreased from 280 GPa to 264 GPa. (E, F) Electrical resistance ($R$) as a function of temperature ($T$) for the sample of (La,Sc)$H_x$ at 264 GPa, measured in a four-probe van der Pauw configuration (channels Ch1-Ch4) using a delta mode with DC current of 0.1 mA. Panel (E) shows channels Ch2 and Ch4, as well as Ch1 and Ch3 in a cooling cycle. Panel (F) shows channels Ch2 and Ch4, as well as Ch1 and Ch3 in a warming cycle.

The second experiment was conducted using an electric DAC SL-2 at a pressure of 256–251 GPa (Fig. 5). The sample preparation, culet size, gasket material, and method for selecting $LaSc_2$ particles—based on pressing a piece of $LaSc_2$ to a thickness of approximately 1 µm—were identical to those used for DAC SL-1. The initial pressure before infrared (IR) laser heating was 256 GPa (Akahama scale (46)). After heating with a 1.06 µm IR laser using a pulse length of 0.05 s, the pressure dropped to 251 GPa (Fig. 5 C, D), and pronounced changes in the morphology and geometry of the sample were observed (Fig. 5 A, B). As in the case of DAC SL-1, no signs of a superconducting transition were detected in electrical resistance measurements using a four-channel van der Pauw scheme over the temperature range of 230 K to 300 K (see Fig. 5 E, F; $I_{DC}$ = 0.1 mA, delta mode (47)). The sample behaved as a typical metal, exhibiting a linear *R(T)* dependence. This DAC broke during the third stage of laser heating.

In the third experiment, we used the same La-Sc alloy with a 1:2 element ratio for the synthesis. The laser heating was carried out at 213 GPa. As the result, the pressure in the DAC SL-3 was increased to 220 GPa and then to 263 GPa (SI Appendix, Fig. S5). Although the data for 213 and 220 GPa were previously published (6), it is interesting to note that along with the compression to 263 GPa, a pronounced trend of decreasing $T_c$ is observed from approximately 241 K at 220 GPa to 195-230 K at 263 GPa ($dT_c/dP < -0.25$ K/GPa), depending on the electrode combination (SI Appendix, Figs. S2 A, B). Apparently, some chemical reaction occurs, and at high pressure, in the presence of high scandium content in the initial alloy, a phase that does not exhibit high-$T_c$ superconductivity is stabilized.

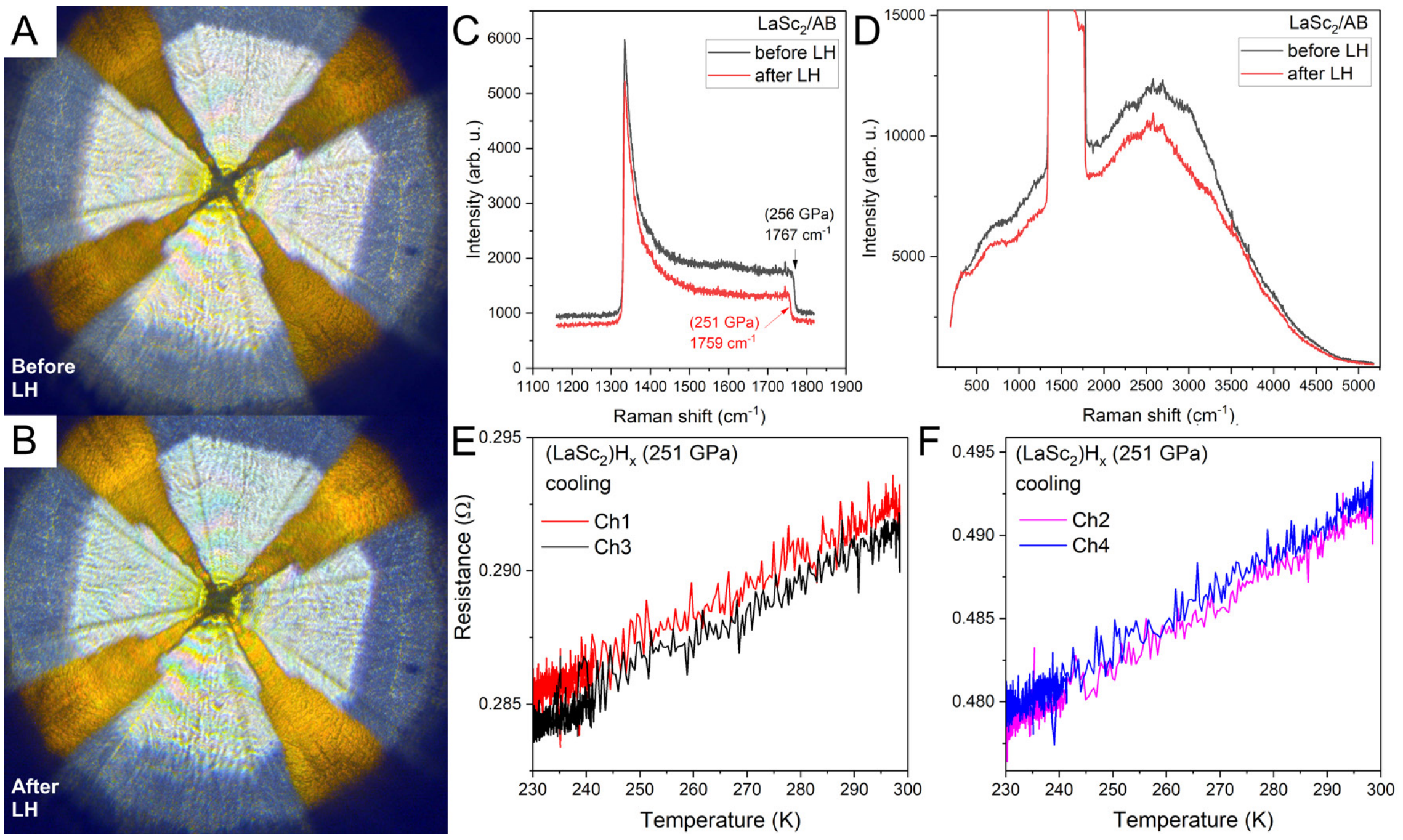

**Fig. 5.** Synthesis of (La,Sc)$H_x$ using infrared pulsed laser heating (LH) under high pressure in DAC SL-2. (a, b) Optical micrographs showing the sample chamber and electrical contacts before (A) and after (B) laser heating at high pressure. After LH we can see a blackening of the sample's surface. (C, D) Raman spectra collected before and after laser synthesis in (C) the region of the diamond Raman edge, and in (D) the full range. During laser heating, pressure in the DAC decreased from 256 GPa to 251 GPa. (E, F) Electrical resistance ($R$) as a function of temperature ($T$) for the sample of (La,Sc)$H_x$ at 251 GPa, measured in a four-probe van der Pauw configuration (channels Ch1-Ch4) using a delta mode with DC current of 0.1 mA. Panel (E) shows channels Ch1 and Ch3, whereas panel (F) shows channels Ch2 and Ch4, both in a cooling cycle.

Experiments in symmetrical Mao-type DACs Z1-Z4 with 20-µm diamond anvils were less successful, yielding only a 25% (1/4) survival rate during compression to ~260 GPa, pulsed laser heating, temperature-dependent electrical resistance studies, or synchrotron X-ray diffraction experiments (SI Appendix, Fig. S6). Only the DAC Z3 sample survived laser heating and demonstrated signs of a superconducting transition, with a maximum $T_c$(onset) of 64 K at 258 GPa (SI Appendix, Fig. S7B-C). This small step in the temperature dependence of electrical resistance $R(T)$ differs significantly from superconductivity in the initial precursor, whose $T_c$ is 3.1–3.2 K at 0 GPa as was found using a SQUID magnetometer (SI Appendix, Fig. S7A). It also cannot be attributed to superconductivity in La or Sc, whose $T_c$ does not exceed 36 K in this pressure range (48). Considering the shift of the transition at 64 K in a magnetic field (0–12 T), we attribute this transition to the manifestation of superconductivity in the unknown hydride $LaSc_2H_x$.

We attempted to study the DAC Z3 sample using synchrotron X-ray diffraction (SI Appendix, Fig. S8). We found that samples with high scandium content (e.g., $ScH_x$) are very poorly visible or not visible at all in the diffraction pattern, whereas 1-µm Pt electrodes are much more visible, and the pressure according to the Pt equation of state (49) corresponds to the pressure according to the Raman signal of diamond.

## Conclusions

Theoretical analysis confirms previous predictions and points to the presence of a sharp maximum of the critical temperature $T_c$ in the La-Sc-H system at the La:Sc ratio of 1:2 for $P6/mmm$-$LaSc_2H_{24}$. The exceptionally strong electron-phonon interaction in this compound may limit the stability of the metallic state of hydrogen sublattice and restrict $T_c$ to values of approximately 270-280 K. The obtained experimental data indicate the absence of near-room temperature superconductivity above 250 K in samples of (La,$Sc_2$)$H_x$ synthesized from the La-Sc (1:2) alloy at pressures of 220 to 264 GPa. A possible reason for the failure of the $P6/mmm$-$LaSc_2H_{24}$ synthesis may be a short total time (0.05-0.3 s) of the laser heating, insufficient to fully order the position of lanthanum and scandium atoms from the disordered solid solution state of the La-Sc alloy. The choice of the starting precursor for the synthesis of $LaSc_2H_{24}$ can play a critical role.

**Materials and Methods**

La–Sc precursor alloys of 1:2 stoichiometric composition were prepared by arc melting of La and Sc metal pieces (total mass approximately 1–2 g) for 1–3 minutes at temperatures above 2000 °C under an argon atmosphere. Prior to loading into diamond anvil cells (DACs), alloy particles were mechanically pressed to a thickness less than 1 μm and studied by scanning electron microscopy (SEM) combined with energy-dispersive X-ray (EDX) analysis to identify homogeneous regions of approximately 10 × 10 μm² exhibiting a La:Sc atomic ratio of 1:2 (≈ 62 at.% Sc). Typically, lanthanum-scandium alloys are very inhomogeneous (SI Appendix, Fig. S1 A). Regions with excessive compositional variation were excluded from use.

High-pressure experiments (DAC SL1-3) were performed using electric DACs equipped with HPHT diamond anvils having culet diameters of 30 μm beveled to 300 μm with an angle of 8.5°. The diamond culets were protected by a sputtered submicron aluminum oxide ($Al_2O_3$) layer to prevent hydrogen cracking. Composite gaskets consisting of W/$CaF_2$/epoxy were prepared for these experiments. For both series of experiments, in DACs SL1-3 and DACs Z1-4, selected La–Sc microparticles were loaded together with ammonia borane ($NH_3BH_3$) used as the hydrogen source. Pressure in DACs was determined by the Akahama pressure scale using the diamond Raman edge. Laser heating was performed using a pulsed infrared laser (wavelength 1.04 μm, a pulse duration was 0.05 s).

DACs Z1-Z4 were equipped with HPHT diamond anvils having culet diameters of 20 μm beveled to 250 μm with an angle of 8.5°. Composite gaskets consisting of W/$Al_2O_3$/epoxy were prepared for these experiments. In this case, protection of the diamond anvils with an $Al_2O_3$ layer was not used, which probably influenced the lower percentage of anvil survival in this series of experiments.

Electrical resistance as a function of temperature was measured in a four-probe van der Pauw configuration using four Au/Ta electrodes. A DC current of 0.1 mA was applied in delta mode to minimize thermoelectric offset errors. Measurements were conducted in both cooling and warming cycles over the temperature range of 230–300 K for SL-1 and SL-2. Four independent measurement channels (Ch1–Ch4) were recorded simultaneously to confirm sample continuity and spatial homogeneity. All DACs, except DAC Z3, were broken during repeated laser heating attempts to see traces of room-temperature superconductivity and $LaSc_2H_{24}$. Due to the rapid (within days) failure of the DACs with hydrogen at 2.5 Mbar and above, X-ray studies were not possible for most of the samples.

X-ray diffraction analysis of the $LaSc_2$ precursor was performed on a Rigaku SmartLab diffractometer with rotating silver anode (λ = 0.56 Å). Synchrotron investigation of the $LaSc_2H_x$ sample in DAC Z3 at 258 GPa was performed on a BL10XU (SPring-8) beamline using an X-ray beam of 30 keV (λ = 0.413 Å) energy and diameters of 10 and 3 microns, yielding identical results. Data acquisition on a Perkin Elmer digital X-ray flat panel detector was 60-120 seconds (2 seconds, 30-60 averages).

Density functional theory (DFT) and density functional perturbation theory (DFPT) calculations were performed using the Quantum Espresso code. Electron-phonon interaction parameters and Eliashberg spectral functions $\alpha^2F(\omega)$ were calculated using the tetrahedron method. Tetrahedron calculations employed a k-point mesh of 12 × 12 × 12, a phonon q-point mesh of 4 × 4 × 4, and the scalar-relativistic PAW PBE pseudopotentials. Electron band structure calculations used a denser k-mesh of 24 × 24 × 24. Calculations were performed for $P6/mmm$-$LaSc_2H_{24}$, $P6/mmm$-$La_3H_{24}$, and $P6/mmm$-$Sc_3H_{24}$ at 300 GPa in the harmonic approximation. Superconducting critical temperatures were estimated using the Allen–Dynes formula with a Coulomb pseudopotential $\mu^* = 0.1$.

**Data, Materials, and Software Availability**

Authors declare that the main data supporting our findings of this study are contained within the paper and Supporting Information. All relevant data are available from the corresponding authors upon request.

**Acknowledgments**

D.V.S. and D.Z. are grateful for the financial support from HPSTAR. This work was supported by the National Key Research and Development Program of China (grant 2023YFA1608900, subproject 2023YFA1608903). V.V.S. acknowledges financial support from the Shanghai Science and Technology Committee, China (No. 22JC1410300) and Shanghai Key Laboratory of Materials Frontier Research in Extreme Environments (MFree), China (No. 22dz2260800). Z. W. thanks Natural Science Foundation China (NSFC), grant no. 52502334 for financial support.

**Contributions**

I.A.T. performed the high-pressure experiments with DACs SL1–SL3. Z. W. independently carried out experiments with DACs Z1-Z4. D.V.S. and D.Z. performed the theoretical calculations, XRD, and a part of transport measurements. E.A.Y. and B.L.A. contributed to the theoretical analysis and interpretation of the results. D.V.S., E.A.Y., B.L.A., and V.V.S. contributed to writing and revising the manuscript. D.V.S. and V.V.S. supervised the project. All authors discussed the results presented in the manuscript and provided comments.

**Significance Statement**

The reproducibility of findings is a key factor when investigating a room-temperature superconductivity in superhydrides. This work focuses on *P6/mmm*-$LaSc_2H_{24}$, which is examined from the perspectives of theoretical prediction and experimental synthesis. We demonstrate that the tetrahedral method confirms a potential room-$T_c$ superconductivity in $LaSc_2H_{24}$ at 300 GPa. Extreme electron-phonon interactions place the metallic state in this hydride at the edge of structural stability, likely capping practical $T_c$ values near 270–280 K at pressures well above 200 GPa. Crucially, seven experimental attempts to synthesize this compound from a $LaSc_2$ alloy failed at 254-280 GPa possibly due to insufficient atomic ordering during pulsed laser heating. These findings pinpoint precursor engineering as the possible key to successful synthesis of ternary superhydrides.

**Conflict of interest**

The authors declare that they have no conflict of interest

# Supporting Information

## Stability Analysis of Superconductivity in *P6/mmm*-$LaSc_2H_{24}$ and its Experimental Reproducibility from La-Sc Alloys

A

B

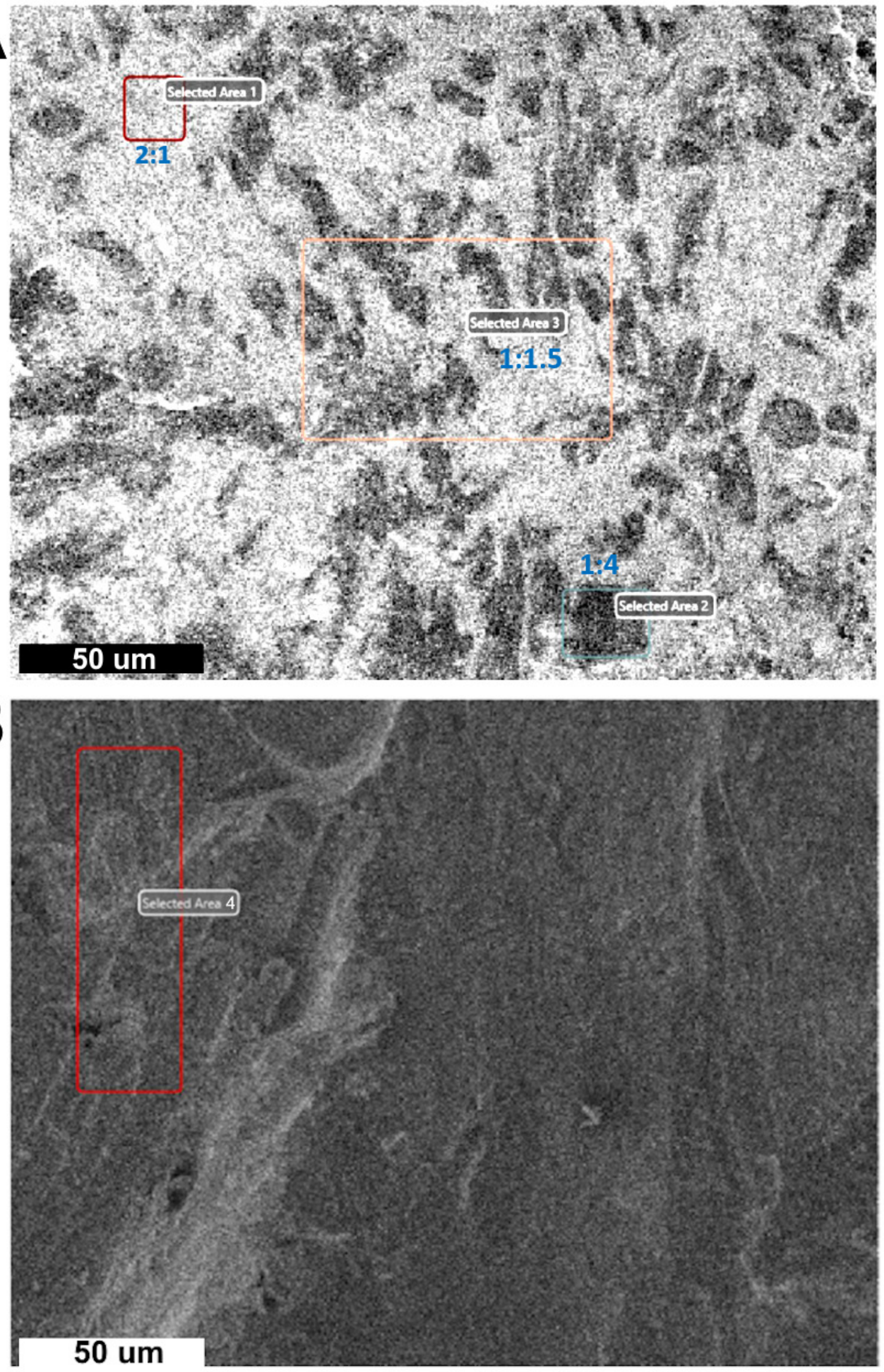

**Selected Area 1**

| Atom | wt% | at% | error% |
|---|---|---|---|
| O | 2.8 | 16.0 | 12.9 |
| Ne | 0.6 | 2.7 | 17.4 |
| Sc | 13.0 | 26.4 | 4.6 |
| La | 82.7 | 54.3 | 3.9 |
| Nd | 0.1 | 0.1 | 78.6 |
| Pm | 0.8 | 0.5 | 60.7 |

**Selected Area 2**

| Atom | wt% | at% | error% |
|---|---|---|---|
| O | 5.4 | 18.4 | 14.5 |
| Sc | 54.0 | 65.7 | 2.9 |
| La | 39.5 | 15.5 | 5.4 |
| Pm | 0.9 | 0.3 | 59.8 |
| Sm | 0.2 | 0.1 | 61.3 |

**Selected Area 3**

| Atom | wt% | at% | error% |
|---|---|---|---|
| O | 4.7 | 20.2 | 13.4 |
| Ne | 0.4 | 1.5 | 20.8 |
| Sc | 29.9 | 46.0 | 3.6 |
| La | 63.8 | 31.8 | 4.4 |
| Nd | 0.0 | 0.0 | 100.0 |
| Pm | 0.7 | 0.3 | 61.1 |

**Selected Area 4**

| Atom | wt% | at% | error% |
|---|---|---|---|
| Al | 0.7 | 1.8 | 18.0 |
| Si | 0.8 | 2.2 | 16.1 |
| Sc | 36.8 | 62.2 | 3.5 |
| La | 61.7 | 33.7 | 4.7 |

**Fig. S1.** Scanning electron microscopy (SEM) images and energy-dispersive X-ray (EDX) elemental analysis of ($La,Sc_2$) alloy samples at different regions. (A) An inhomogeneous La–Sc alloy sample exhibiting a 3–4-fold variation in Sc concentration across a region of approximately 50 µm, as evidenced by EDX measurements in three selected areas (La:Sc atomic ratios of approximately 2:1, 1:1.5, and 1:4). (B) A more homogeneous ($La,Sc_2$) alloy sample used for loading DACs SL-1, 2, and 3, showing a uniform Sc concentration (≈ 62 at.%) across the analyzed region.

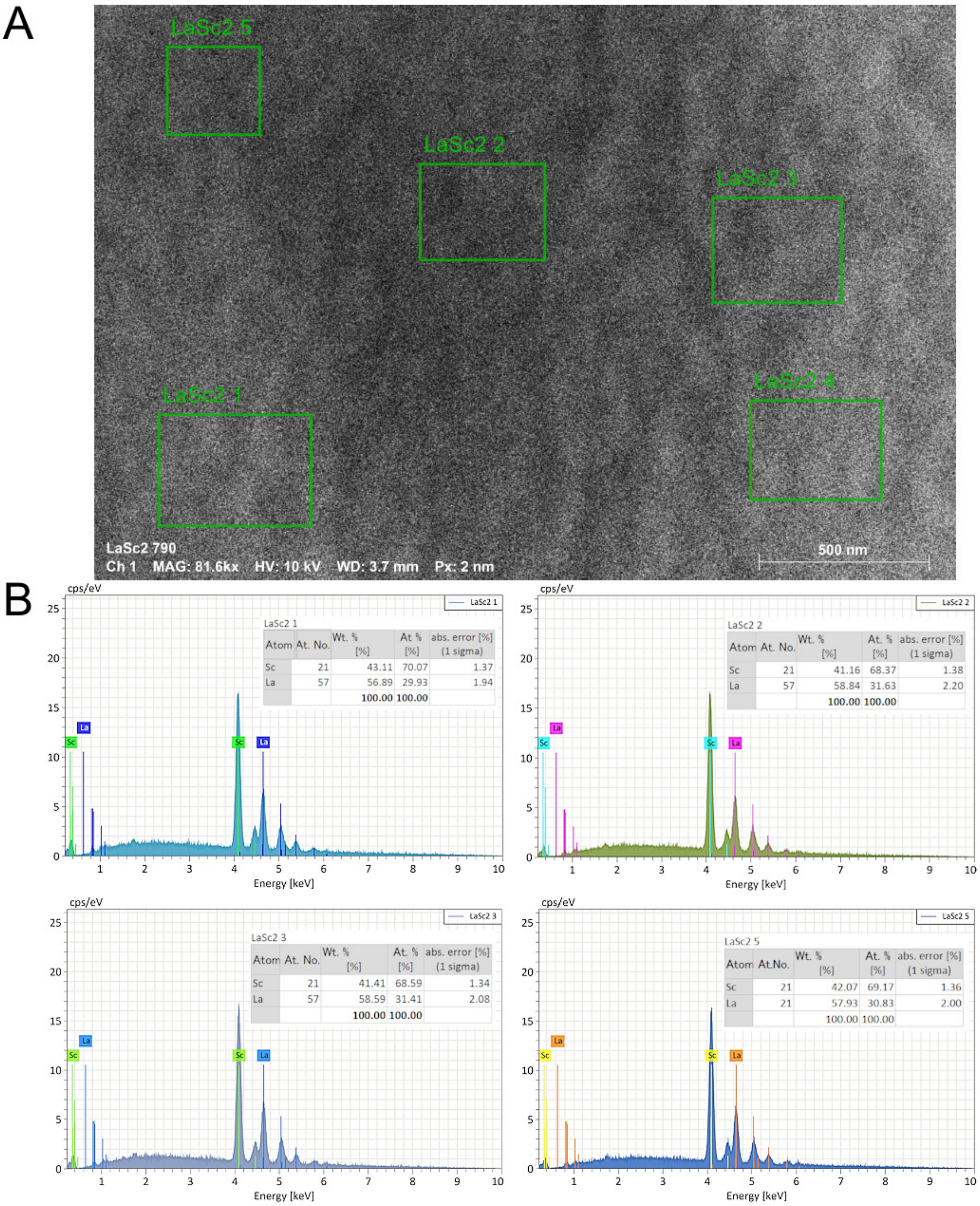

**Fig. S2.** Scanning electron microscopy image and energy-dispersive X-ray elemental analysis of (La,$Sc_2$) alloy used for DACs Z1-Z4 loading at different regions. (A) A homogeneous La–Sc alloy sample analyzed in four selected areas. (B) EDX analysis and concentration of elements in the selected areas. Variation of the Sc concertation is about ± 1-2 at. %.

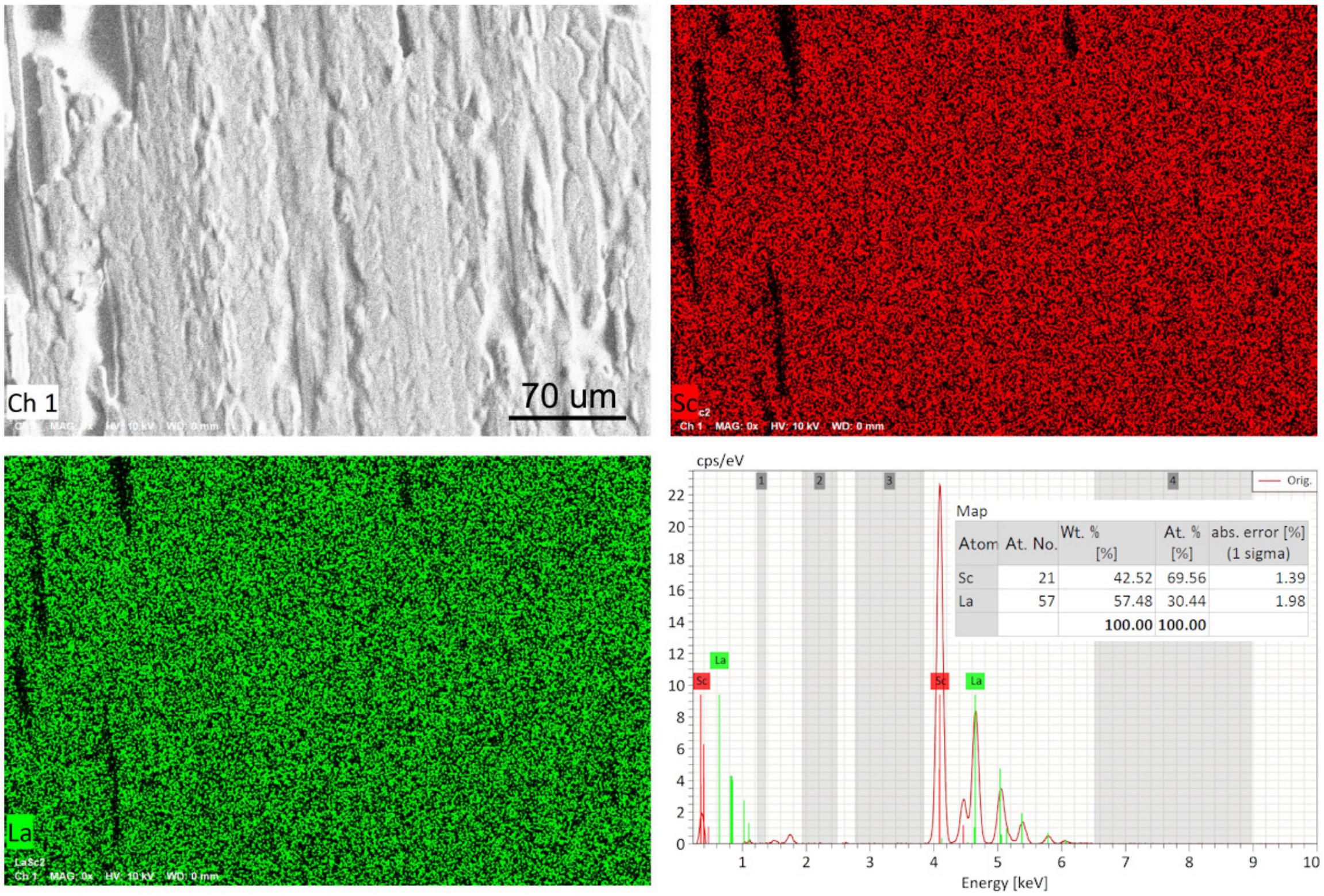

**Fig. S3.** Scanning electron microscopy large-scale image, EDX analysis and mapping of the elements (La, Sc) distribution for the homogeneous (La,$Sc_2$) alloy used for DACs Z1-Z4 loading. Variation of the Sc concertation is about ± 1-2 at. %.

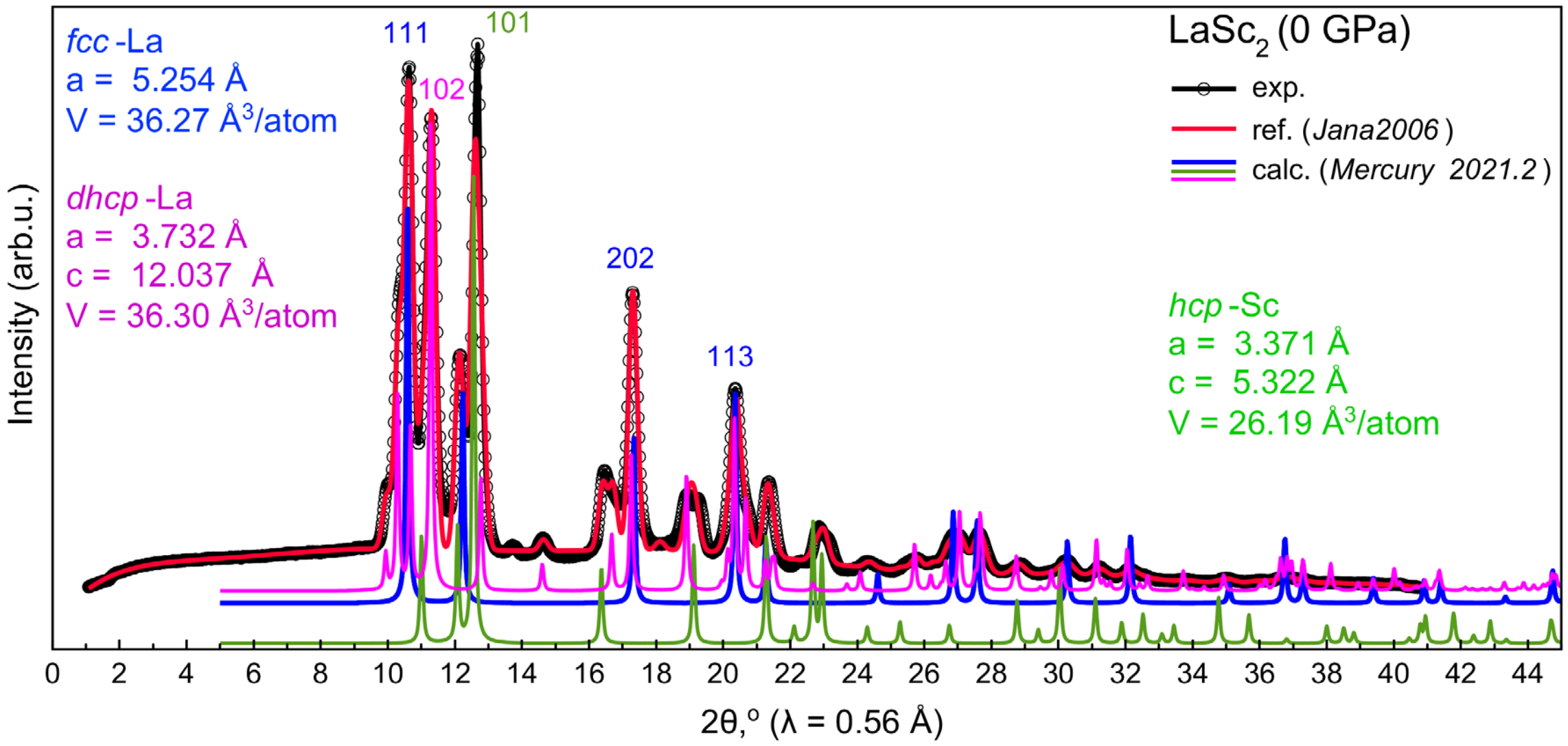

**Fig. S4.** X-ray diffraction pattern (Ag anode) of the initial $LaSc_2$ alloy at ambient pressure. Black circles are the experimental data; the red line is a Le Bail refinement combining three reference phases: *fcc*-La (blue, a = 5.254 Å, V = 36.27 Å$^3$/atom), *dhcp*-La (magenta, a = 3.732 Å, c = 12.037 Å, V = 36.30 Å$^3$/atom), and *hcp*-Sc (olive, a = 3.371 Å, c = 5.322 Å, V = 26.19 Å$^3$/atom). XRD patterns were predicted with Mercury 2021.2. A few the most intense reflections were indexed.

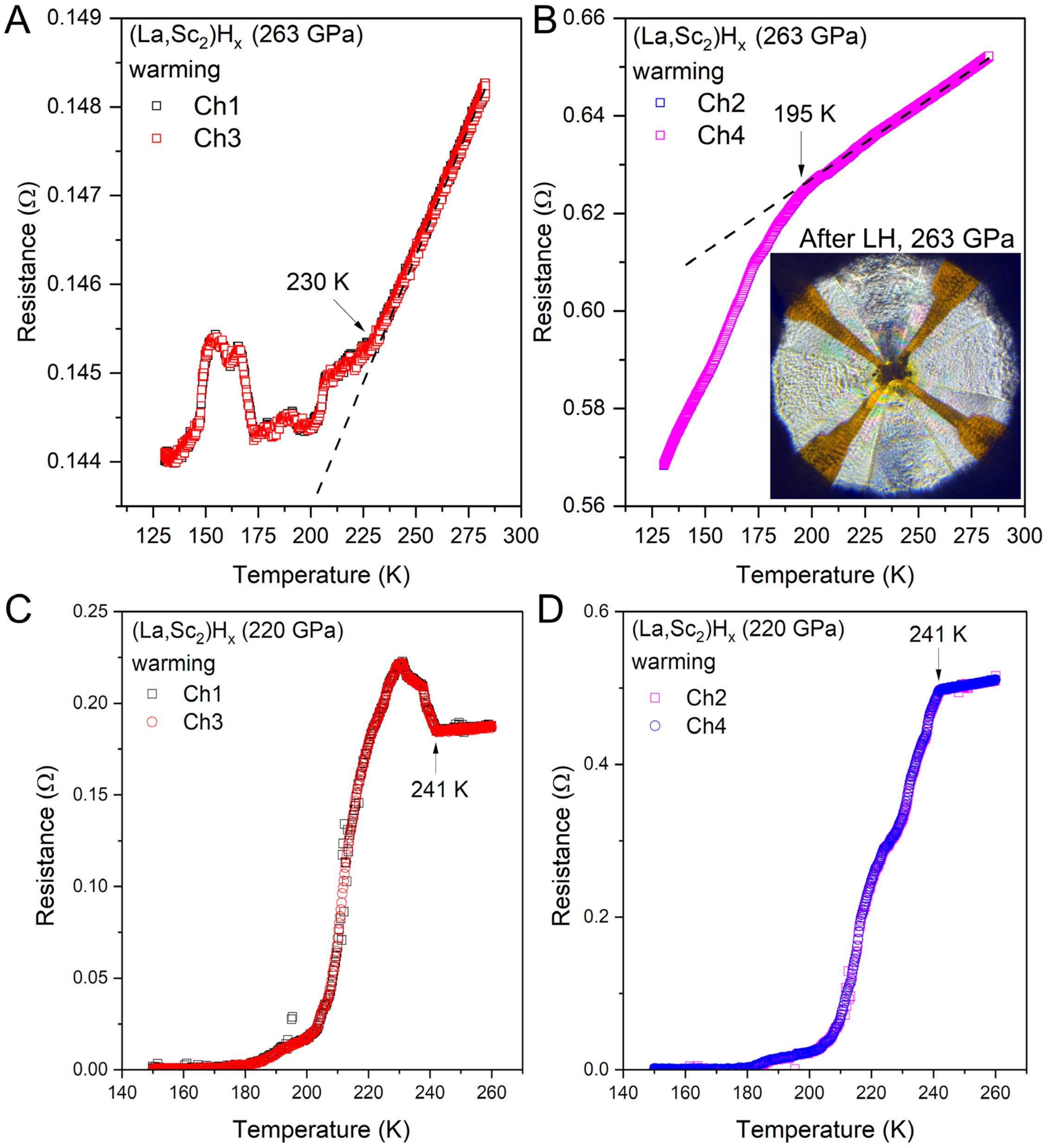

**Fig. S5.** Temperature-dependent electrical resistance measurements of $(La,Sc_2)H_x$ hydride in DAC SL-3 at 220-263 GPa. (A) Resistance vs. temperature for two van der Pauw channels (Ch1 and Ch3) at 263 GPa, showing usual metallic behavior of the sample above 230 K. The deviation from the dashed line indicates possible onset of the superconducting transition. (B) Electrical resistance measurements for channels Ch2 and Ch4 at 263 GPa, displaying metallic behavior ($dR/dT > 0$) with a kink feature at approximately 195 K. Inset: Optical microscope image of the sample after laser heating (LH) at 263 GPa, showing the diamond anvil cell culet with the sample chamber and electrodes. (C) Resistance vs. temperature at 220 GPa for channels Ch1 and Ch3 recorded during warming cycle, exhibiting a pronounced superconducting transition with resistance dropping to near-zero below 241 K (onset). (D) Resistance measurements for channels Ch2 and Ch4 at 220 GPa, showing a clear superconducting transition starting at 241 K with resistance dropping from ~0.5 Ω to near-zero values.

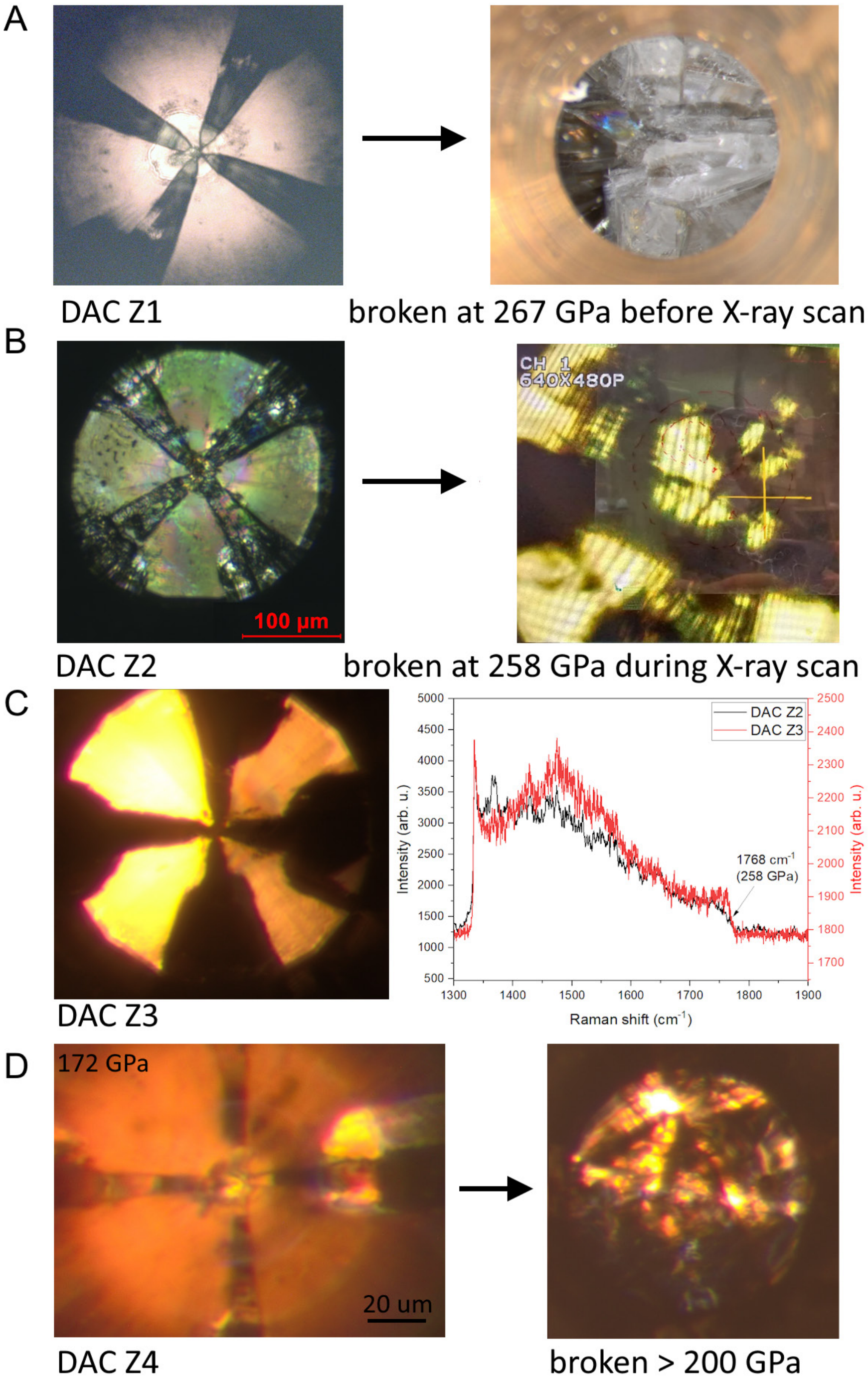

**Fig. S6.** Optical photos and high-pressure Raman spectra of La-Sc (1:2) alloy or corresponding hydride samples in DACs Z1–Z4. (A) DAC Z1**:** optical image of the sample chamber showing four Pt electrical contacts on the diamond culet at 267 GPa, and the fractured diamond anvil after cell failure prior to synchrotron X-ray diffraction. (B) DAC Z2: photo of the four-probe Pt electrode system sputtered on the diamond culet, and the broken anvil at 258 GPa observed after an X-ray mapping using photo diode. (C) DAC Z3: optical photo of the pressurized $(La,Sc_2)H_x$ sample and Pt electrodes in transmitted light, alongside with Raman spectra collected for DACs Z2 and Z3. (D) DAC Z4: photo of the sample chamber loaded with $LaSc_2$ alloy at 172 GPa, and the damaged diamond culet photo following anvils failure at a pressure exceeding 200 GPa.

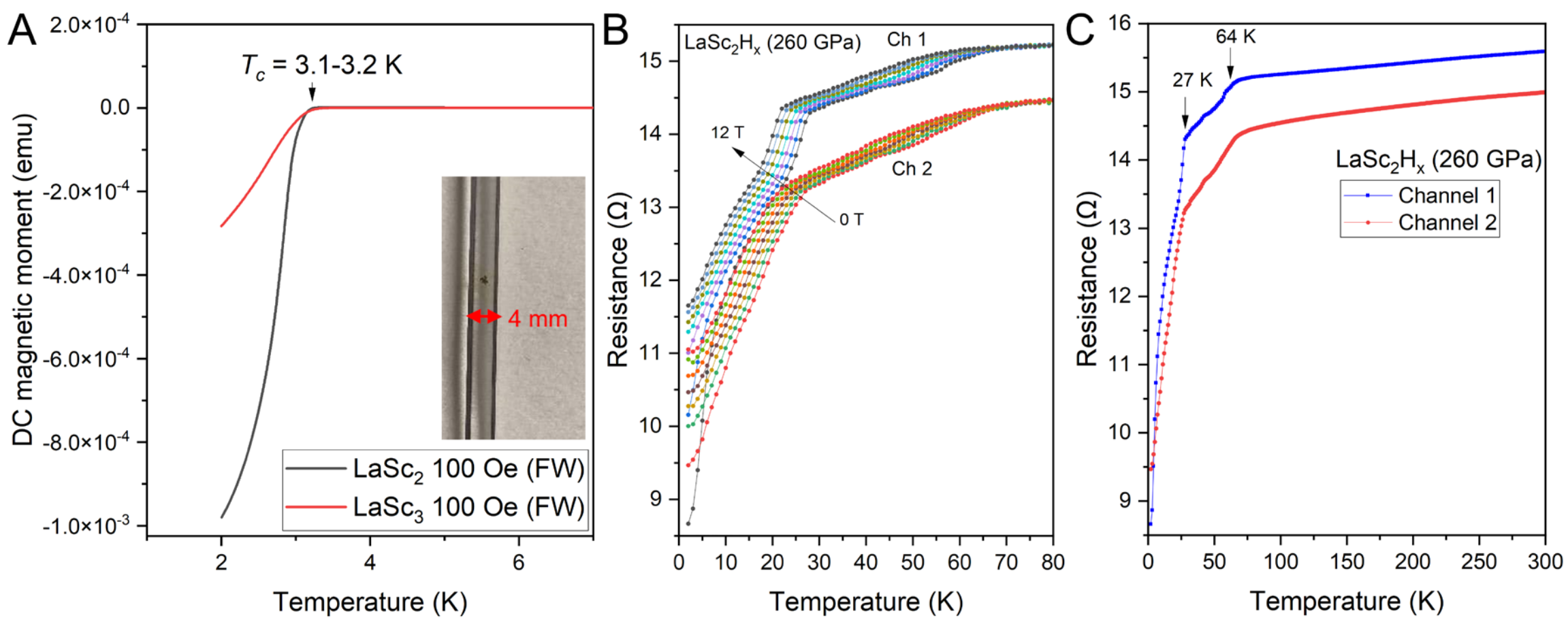

**Fig. S7.** Superconducting properties of La-Sc alloys at ambient pressure, and $LaSc_2H_x$ hydride synthesized from $LaSc_2$ alloy at 258 GPa in DAC Z3. (A) Temperature dependence of magnetic moment of the $LaSc_2$ (black curve) and $LaSc_3$ (red curve) alloys measured using a SQUID VSM magnetometer (Quantum Design). Both alloys exhibit a $T_c$ around 3.1–3.2 K. Inset: a particle of $LaSc_2$ alloy glued to the $SiO_2$ probe for the SQUID measurements. (B) Superconducting transitions in DAC Z3 at 258 GPa and their shift to lower temperatures when applying a magnetic field from 0 to 12 T. (C) Temperature dependence of electrical resistance of DAC Z3 over a wide temperature range. The highest onset $T_c$ corresponds to 64 K.

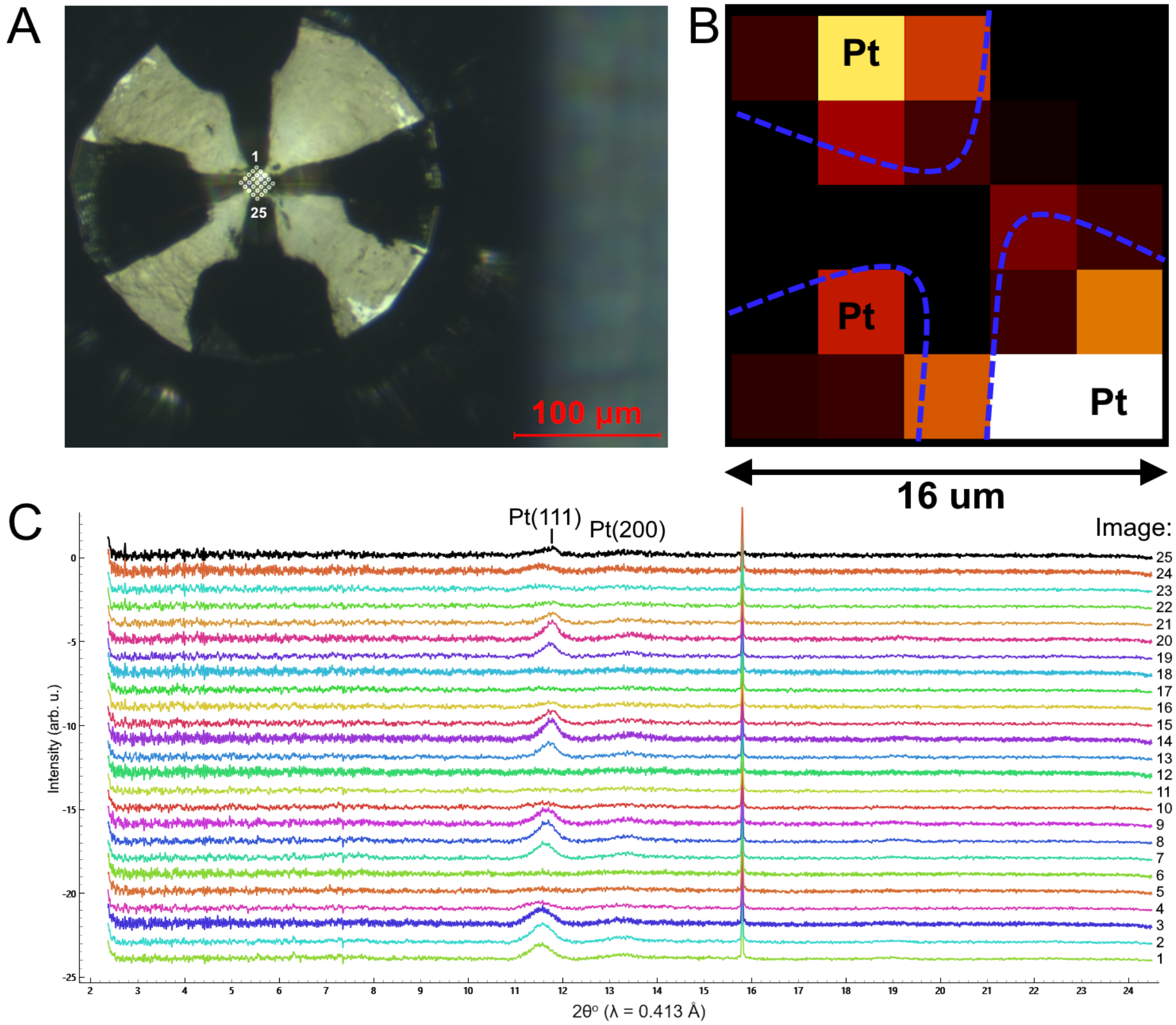

**Fig. S8.** A series of X-ray diffraction patterns and corresponding XRD map of the DAC Z3 sample at 258 GPa (300 K). Wavelength is 0.413 Å. (A) Overlay of the diffraction map (16 x 16 μm$^2$) on the optical photo of the sample. (B) XRD intensity map of the platinum electrodes (Pt was sputtered by Xe FIB). (C) Stack of integrated diffraction patterns, showing only the Pt electrodes broad reflections. The initial thickness Pt electrodes was 1 μm. The maximum pressure according to the equation of state of Pt is about 260 GPa.